# Generating High-Precision Force Fields for Molecular Dynamics Simulations to Study Chemical Reaction Mechanisms using Molecular Configuration Transformer

Sihao Yuan[1,2], Xu Han[1], Jun Zhang[3], Zhaoxin Xie[1,2], Cheng Fan[1,2], Yunlong Xiao[1], Yi Qin Gao[1,2,3,*], and Yi Isaac Yang[2,*]

1  Institute of Theoretical and Computational Chemistry, College of Chemistry and Molecular Engineering, Peking University, Beijing 100871, China
2  Institute of Systems and Physical Biology, Shenzhen Bay Laboratory, Shenzhen 518132, China
3  Changping Laboratory, Beijing 102200, China
*  To whom correspondence should be addressed: yangyi@szbl.ac.cn or gaoyq@pku.edu.cn

## Abstract

Theoretical studies on chemical reaction mechanisms have been crucial in organic chemistry. Traditionally, calculating the manually constructed molecular conformations of transition states for chemical reactions using quantum chemical calculations is the most commonly used method. However, this way is heavily dependent on individual experience and chemical intuition. In our previous study, we proposed a research paradigm that uses enhanced sampling in molecular dynamics simulations to study chemical reactions. This approach can directly simulate the entire process of a chemical reaction. However, the computational speed limits the use of high-precision potential energy functions for simulations. To address this issue, we present a scheme for training high-precision force fields for molecular modeling using a previously developed graph-neural-network-based molecular model, molecular configuration transformer. This potential energy function allows for highly accurate simulations at a low computational cost, leading to more precise calculations of the mechanism of chemical reactions. We applied this approach to study a Claisen rearrangement reaction and a Carbonyl insertion reaction catalyzed by Manganese.

# Introduction

The theoretical research of chemical reactions is crucial in chemistry[1] and biology[2]. We can study the mechanism of chemical reactions by solving many-electron Schrödinger equations[3]. These quantum chemical calculations can provide detailed potential energy surfaces and the corresponding electronic properties of molecular configurations. Generally, to illustrate the reaction mechanism of a chemical reaction, many possible reaction pathways should be explored, which quite often requires experimental information[1] and initial guess[4] about structures of intermediates and transition states. The proposed pathways heavily depend on individual experience and chemical intuition when experimental evidence is lacking. Several algorithms exist to locate transition states in chemical systems, including nudged-elastic-band method[5,6], string method[7,8], partitioned rational-function optimizer[9–11], dimer method[12–14], among others. However, as the complexity of reaction systems increases, an increasing number of local minima in the configuration space and entropy effects from energetically similar configurations become more significant[15,16]. Traditional methods based on single-point energy calculations are difficult to traverse the entire configuration space effectively and may miss low-energy conformations[17], resulting in wrong understandings of reactions.

Molecular dynamics (MD) simulation is an effective research tool for complex chemical reaction systems. Compared to chemical reaction studies based on single-point energy calculations, MD simulations can directly consider temperature,[18] entropy,[19,20] as well as solvent effects[16,21], and can be used to calculate the reaction free energy[22,23] and kinetics[24,25]. They can also explore reaction mechanisms independent of initial assumptions about reaction pathways[16,26]. However, it is extremely difficult for individual molecules in MD simulations to spontaneously transform into transition states because of the high free energy barriers of chemical reactions. Therefore, studying chemical reactions using MD simulations usually requires enhanced sampling methods[27].

Enhanced sampling methods can substantially increase the sampling efficiency of rare events in MD simulations, to allow the observation of transition state conformations, thereby facilitating chemical reactions. Many enhanced sampling methods are based on collective variables (CVs)[28]. For example, the metadynamics (MetaD)[28,29] method is a CV-based enhanced sampling method commonly used for chemical reactions. MetaD can continuously accumulate Gaussian forms of bias potential in the degrees of freedom (DOFs) associated with CVs to facilitate the exploration of the system over the space of CVs, thus advancing the transformation between reactants and products. MetaD has been successfully applied to many chemical reactions such as reduction of the Pt(IV) Asplatin Prodrug[30], the azide–alkyne Huisgen cycloaddition[31], and was used to study the selectivity in the Wittig reaction[32]. There are also enhanced sampling methods without requiring CVs, which can also be used to study chemical reactions. For example, the ITS-enhanced sampling method[33,34] has been successfully applied to study slow processes and chemical reactions such as β-structures folding of peptides[35,36], Claisen rearrangement of allyl vinyl ether[16,37], and salt-water clusters formation[38–40]. In addition, the CV-free ITS method can be combined with the CV-based MetaD method, as in the multi-scale enhanced sampling method MetaITS[41,42], to achieve more efficient sampling of the simulation system.

The efficiency of the CV-based enhanced sampling methods significantly relies on the quality of the defined CVs, and obtaining efficient CVs for complex systems is challenging. As a result, some optimization methods for CVs have been developed, including Harmonic Linear Discriminant Analysis (HLDA) [43,44], time-lagged component analysis (TICA)[45,46], Skewed-Momenta method[47] and spectral gap optimization method for order parameters (SGOOP)[48]. Among them, HLDA is commonly used for chemical reactions. HLDA can generate CVs through linear combinations of provided descriptors without necessitating information about transition states[30,31].

The potential energy function or force field is another crucial factor in MD simulations of chemical reactions. Commonly used force fields and applicable to chemical reactions include *ab initio* methods, semi-empirical methods[49–51], reactive force fields, and machine learning force fields. The *ab initio* molecular dynamics (AIMD) simulation using quantum mechanics (QM) calculations is the most accurate method. However, it is computationally costly and is usually only used for short time simulations of small systems. The semi-empirical methods use empirical parameters instead of computationally expensive integrals to calculate electronic states. Therefore, they are faster, but often result in accuracy inferior to AIMD methods. The reactive force field methods[52,53] do not compute electronic states but directly employ functions of atomic positions as energy, so they are faster but less accurate. With the rapid development of Artificial Intelligence (AI), the application of machine-learning based force fields became popular in recent years. In particular, the recent development of those deep neural network-based force fields, such as BPNN[54], DPMD[55], EANN[56], SchNet[57] and PhysNet[58] is expected to help achieve both high speed and accuracy[59,60].

Here, we present a strategy for generating high-precision force fields for chemical reactions. This force field is based on the MolCT model, a GNN-based deep molecular model developed in our group. We use the MolCT model to fit high-precision QM calculation results to achieve an accurate and efficient force field. To efficiently obtain the training dataset for the force field, we also designed a multi-scale sampling and computational scheme: implementing enhanced sampling for the chemical reaction in MD simulations with low-precision force fields and then selecting conformations from the sampling for high-precision QM calculations. We used this method to generate force fields and run MD simulations for the Claisen rearrangement reaction and the manganese-catalyzed carbonyl insertion reaction.

## Methodology

### A. Molecular configuration transformer

Molecular Configuration Transformer (MolCT)[61] is a deep molecular model based on the graph neural network (GNN)[62]. A deep molecular model is a neural network encoder that encodes high-dimensional molecular information, including conformation and environment into low-dimensional vectors or scalars. It can be used to fit molecular properties such as energy and force. For GNN-based deep molecular models, it treats the molecular system as a graph consisting of nodes (or vertices) and edges. The model encodes the atoms in the molecule and its environment as node vectors $\{n_i\}$, and the information between the atoms (e.g., distances) as edge vectors $\{v_{ij}\}$. These node and edge vectors are encoded as the representation vectors of the molecule after several iterations using the algorithm of the interaction layer, which is the core algorithm of the GNN-based deep molecular model. The interaction layer of the MolCT model uses a so-called "ego-attention" mechanism, which is borrowed from the famous Transformer model[63].

$$n_i^{l+1} = \text{FFN}\left\{n_i^l + \text{LayerNorm}\left[\text{Attention}(Q_i, K_i, V_i)^T\right]\right\} \quad (1)$$

$$\text{Attention}(Q, K, V) = \text{Softmax}\left(\frac{QK^T}{\sqrt{D}}\right)V \quad (2)$$

where "FFN" is a feed-forward network and "LayerNorm" means a layer normalization operator. $Q$, $K$ and $V$ are the query, key and value vectors, respectively:

$$Q_i = n_i \quad (3)$$

$$K_{ij} = \{k_{j|i}\} = \{n_j + v_{ij}W^{(K)}\} \quad (4)$$

$$V_{ij} = \{v_{j|i}\} = \{n_j + v_{ij}W^{(V)}\} \quad (5)$$

where $W^{(K)}$ and $W^{(V)}$ are optimizable affine matrices acting on the edge vectors $\{v_{ij}\}$.

Various properties of molecular systems can be obtained by processing the representation vectors with different readout functions. Here, we use the MolCT model to fit the potential energy surface using data obtained from the QM calculation. The potential energy of a molecular system is decomposed into contributions from individual atoms in the system. We use a simple multilayer perceptron as a readout function to process the individual representation node vectors $n_i^{\text{rep}}$ thereby predicting the energy $E_i$ of each atom $i$ and summing the energies $\{E_i\}$ of all atoms as a prediction of the total energy $E$ of the system. The atomic forces $\{F_i\}$ of the system can be obtained by calculating the negative gradient of the potential energy $E$ with respect to the atomic coordinates $\{R_i\}$ by automatic differentiation of AI framework. When training the model, both the energy $E^{\text{label}}$ and force $\{F_i^{\text{label}}\}$ corresponding to the system need to be added to the dataset as labels, and the loss function is:

$$\text{loss} = w_E L_E\left(E^{\text{pred}}, E^{\text{label}}\right) + w_F \sum_i L_F\left(F_i^{\text{pred}}, F_i^{\text{label}}\right) \tag{6}$$

where $L_E(x, y)$ and $L_F(x, y)$ are the loss function for energy and force, respectively, and $w_E$ and $w_F$ are two hyperparameters that balance energy and force training. See Chapter-S-I in SI for detailed information.

### B. Enhanced sampling methods

The training dataset for the force field needs to contain sufficiently rich representative molecular conformations, thus requiring the employment of enhanced sampling methods. We have developed a well-established protocol in our previous work[64] for enhanced sampling of chemical reactions. First, we need to use some CV-optimization methods to find appropriate CVs. The HLDA method is one of the commonly used methods for chemical reactions, which presents a linear combination of multiple descriptors $\{d_i(R)\}$ in the system with different weights $\{\lambda_i\}$ as CVs: $s_{\text{HLDA}}(R) = \sum_i \lambda_i c_i d_i(R)$. The descriptors are physical quantities that are related to the reaction processes, such as interatomic distances and dihedral angles. See Chapter-II in SI for detailed information

Next, we implement the metadynamics (MetaD) method in MD simulation for enhanced sampling with the optimized-CVs. MetaD achieves enhanced sampling by superimposing Gaussian-type repulsive potentials $G(s, s')$ on the visited CVs $s(t)$ as bias potential $V(s, t)$, thereby smoothing out the energy differences between potential wells and barriers:

$$V(s, t) = \int_0^t d\tau G[s, s(\tau)] = \int_0^t d\tau \omega \exp\left(-\frac{\|s - s(\tau)\|^2}{2\sigma}\right) \tag{7}$$

where $\omega$ is the height of Gaussian.

We can also introduce CV-free ITS methods to further increase the sampling efficiency further. The ITS approach uses an effective potential $U_{\text{eff}}[U(R)]$ instead of the potential energy $U(R)$ of the simulation system, which allows the MD simulation to be equivalent to sampling under a series of summed ensemble distributions at different temperatures $\{\beta_k\}$:

$$U_{\text{eff}}[U(\boldsymbol{R})] = -\frac{1}{\beta_0}\log\sum_k n_k e^{-\beta_k U(\boldsymbol{R})} \quad (8)$$

where $\beta_0 = 1/k_B T$ is the simulation temperature, $n_k$ is the weights for each temperature $\beta_k$. The ITS method can be combined with the MetaD method to achieve a multi-scale enhanced sampling:

$$U'_{\text{eff}}(R) = -\frac{1}{\beta_0}\log\sum_k n_k e^{-\beta_k\{U(\boldsymbol{R})+V[\boldsymbol{s}(\boldsymbol{R})]\}} \quad (9)$$

where $V[\boldsymbol{s}(\boldsymbol{R})]$ is the bias potential of MetaD. Compared to the stand-alone MetaD method, this hybrid enhanced sampling method, which we call MetaITS, can simultaneously enhance the sampling of DOFs in addition to CVs, thus significantly improving sampling efficiency[41,42]. In addition, some restraining bias potentials can also be added to the system to avoid side reactions if necessary. More details can be found in Ref. [64].

### C. Multi-scale simulation and sampling

The usage of effective enhanced sampling methods facilitates the generation of the training datasets for chemical reaction force fields by QM calculations. However, since direct high-precision *ab initio* molecular dynamics (AIMD) simulations are computationally too expensive, we use here a multi-scale simulation and sampling strategy. Firstly, we use semi-empirical calculation methods, e.g. DFTB[65–67], AM1[51], PM6[68], in MD simulations for enhanced sampling of chemical reactions. Although they are less accurate these methods are computationally faster and can be used to quickly generate approximate molecular conformation ensembles.

After sufficient sampling in the phase space of reactions, we proceeded to select representative conformations. Since conformations of the transition states are critical for a successful study of chemical reactions, enough conformations near the transition state should be selected from the simulation trajectories. In the case that the transition state conformation is still insufficient from these enhanced sampling simulations, a shooting strategy can also be used. In this latter approach, trajectories are shot starting from molecular configurations representing the transition state with randomly assigned initial momenta. In addition, repetitive sampling of highly similar conformations of stable states is avoided. These criteria for data selection also apply in data supplementation for the construct of stable neural network force fields as discussed in the final section. Next, we perform high-precision QM calculations on these selected representative molecular conformations. Each conformation's single-point energy and energy gradients are computed, thus making a dataset containing atomic coordinates, potential energies, and atomic forces.

### D. Force field training

The workflow of force field training is shown in Figure 1. Firstly, we perform the multi-scale simulation and sampling strategy as we introduced, thereby making up the training dataset. Then, we use this dataset to train the MolCT model to obtain a highly accurate force field for chemical reactions in MD simulations. When MD simulations are performed on ML-based force fields, the system might reach the un-sampled regions on the PES, even with the usage of enhanced sampling techniques for the generation of the training set. In this case, it is necessary to retrain the force field by supplementing unsampled conformations encountered during the simulation into the dataset.

To efficiently supplement the data for the force field training, it is desired to have the following: 1, The added molecular conformations are chemically reasonable to avoid the contamination of the training dataset by unphysical conformations; and 2, The conformations of the newly added molecules are sufficiently different from these existing in the training dataset. However, for conformational sampling of complex systems, it is very difficult to satisfy such conditions through human experience. To overcome these problems, we proposed a GNN-based deep molecular model as a discriminator $D(\mathbf{R})$ to measure the conformational similarity between molecular conformations in the training dataset. We use an adversarial network[69] approach to train an optimizable discriminator $D_\theta(\mathbf{R})$ with parameters $\{\theta\}$, and using two cross-entropies acting on the positive and negative samples as the loss functions:

$$L = L_{\text{CrossEntropy}}[D_\theta(\mathbf{R}_{\text{pos}}), y_{\text{pos}}] + L_{\text{CrossEntropy}}[D_\theta(\mathbf{R}_{\text{neg}}), y_{\text{neg}}] \tag{10}$$

$$L = -\frac{1}{N_{pos}} \sum_{N_{pos}} \log D_\theta(\mathbf{R}_{pos}) - \frac{1}{N_{neg}} \sum_{N_{neg}} \log[1 - D_\theta(\mathbf{R}_{neg})] \tag{11}$$

It he above equations, $\mathbf{R}_{\text{pos}}$ and $\mathbf{R}_{\text{neg}}$ are the conformations of the positive and negative samples, respectively, while $y_{\text{pos}} = 1$ and $y_{\text{neg}} = 0$ are the labels of the positive and negative samples, respectively. In the training of $D_\theta(\mathbf{R})$, conformations from the training dataset of the force field are used as positive samples $\{\mathbf{R}_{\text{pos}}\}$, while conformations obtained from MD simulations using the MolCT-based force field (including those irrational ones) is used as negative samples $\{\mathbf{R}_{\text{neg}}\}$.

We use the trained discriminator $D_\theta(\mathbf{R})$ to determine which conformations in the MD simulated trajectories can be used as a complement to the training dataset. The conformations with $D_\theta(\mathbf{R}) = 1$ are those that are identical to the conformations in the dataset, while those with $D_\theta(\mathbf{R}) = 0$ are the completely irrational conformations. Neither type of conformation is suitable for addition to the dataset. Generally, we set two thresholds $\gamma_{\max}$ and $\gamma_{\min}$ and select those conformations with $\gamma_{\min} < D_\theta(\mathbf{R}) < \gamma_{\max}$ as supplementary conformations. These conformations show similarity to those in the training dataset, suggesting that they are reasonable chemical structures, but as the same time sufficiently different from those already contained in the dataset. The energies and atomic forces on these conformations will then be calculated using high-precision QM methods and added to the dataset to retrain the MolCT-based force field. MD simulations using the MolCT force fields retrained after supplementing the dataset normally were shown to indeed become more stable. This retraining procedure of the force field with additional data is repeated until the MD simulation using the force field becomes stable. In practice, only one or two data supplement steps are needed.

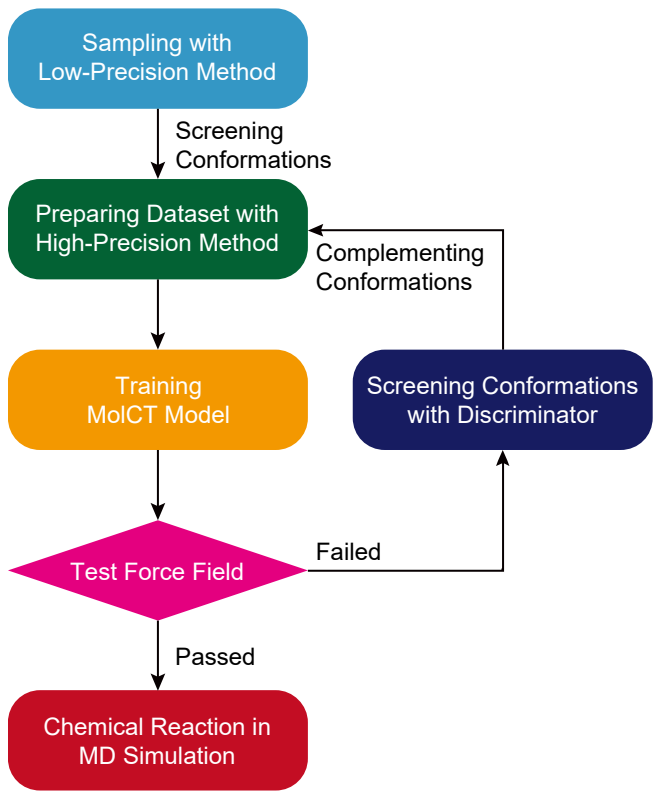

Figure 1. Flowchart of the training process for the MolCT force field.

### E. Software and codes

For the force field training and MD simulations, readers can choose different MD simulation, quantum chemistry, and enhanced sampling software or libraries according to their specific needs.

1) To sample molecular conformations based on low-precision force fields, one can employ MD simulation software equipped with semi-empirical methods (e.g., AMBER[70], CHARMM[71,72], and CP2K[73]). Alternatively, one can also combine MD simulation engines with quantum chemistry software. In this direction, we have developed MindSPONGE[74,75], an MD simulation software that seamlessly integrates with the quantum chemistry software PySCF[76].

2) Enhanced sampling techniques can be integrated into MD simulation software by utilizing specialized libraries like PLUMED[77–79] and COLVARS[80], which support the MetaD method. Notably, our MD simulation software, MindSPONGE, includes built-in enhanced sampling methods such as MetaD, ITS, and MetaITS. Additionally, we offer a code for the CV-optimization method HLDA, which is available in edited PLUMED code (github.com/helloyesterday/plumed2_HLDA), as well as a standalone Python code (https://gitee.com/yuansshh/mdtools).

3) For high-precision quantum mechanical (QM) calculations of molecular conformations, researchers can utilize quantum chemistry software such as Gaussian[81], ORCA[82–84], and PySCF.

4) We have developed an in-house generalized framework Cybertron (https://gitee.com/helloyesterday/cybertron) for GNN-based deep molecular models incorporating MolCT. Cybertron has a built-in MolCT model, which can be used to train MolCT-based force fields. Additionally, Cybertron features an integrated adversarial discriminator training tool, which supplements data for molecular conformations.

5) The MolCT-based force field models trained with Cybertron can be directly employed as potential functions in MindSPONGE for MD simulations. By leveraging enhanced sampling methods, such as MetaD, integrated into MindSPONGE, one can achieve MD simulations of chemical reaction processes.

## Results and Discussion

### Claisen rearrangement

We first test our method on a classical intramolecular Claisen rearrangement of cis-2-vinylcyclopropane carbaldehyde. The schematic diagram presented in Figure 2a depicts the reaction system. The reaction involves the conversion between a seven-membered ring (referred to as molecule A) and a three-membered ring (referred to as molecule B). We first performed MD simulations at 800 K using the AMBER20 software package[70] with the semi-empirical DFTB[65–67] method. The multiple walker metadynamics (MW-MetaD)[85] method was used for enhanced sampling with PLUMED2[78] plugin library. Since the mechanism for this reaction is well understood and straightforward, we simply define the CV simply as the difference between bond formation and bond breaking distances, represented as $d_1 - d_2$. In addition, to avoid undesired byproducts, some additional restraints were added to the system (For more details, please refer to Chapter-S-III in Supporting Information). We performed four parallel MD simulations, each of 10ns long.

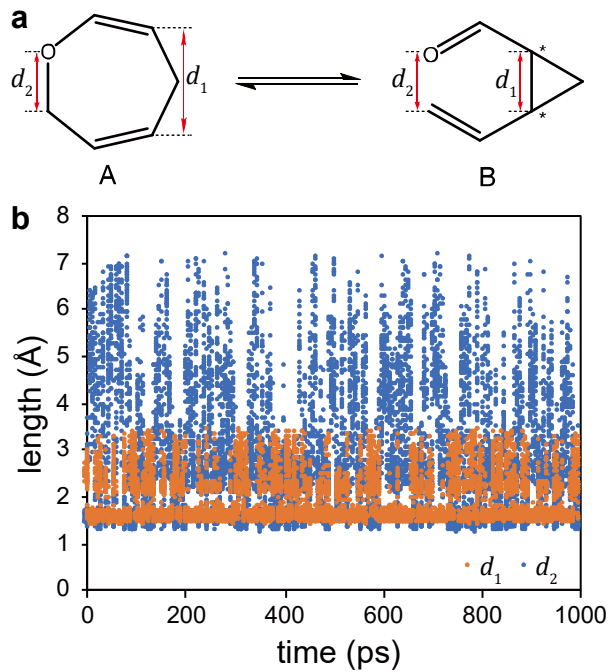

Figure 2. a) The illustration of Claisen rearrangement and the definition of distances $d_1$ & $d_2$. b) The evolution of CV with time in MD simulations with the DFTB method as the force field.

The effect of the sampling is shown in Figure 2, where the molecule has been transformed between reactants and products multiple times. These transitions allow us to extract the molecular conformations for the training dataset from the simulation trajectories. To ensure a sufficient sampling of molecular conformations of the transition states, we retained all recorded configurations near transition states with CVs ranging from -0.5Å to 0.5Å. Other configurations were uniformly sampled along the CV so that a total of 133,120 samples were selected. Subsequently, the single-point energies and energy gradients for these conformations were obtained by performing QM calculations at B3LYP[86–89]/6-31+G** level. The energies and atomic forces from QM calculations, together with the atomic coordinates of the molecules, form the training samples. We partitioned them by random selection into a training set of 131,072 samples, a validation set of 1,024 samples, and a test set of 1,024 samples.

We used this dataset to train the MolCT model, which has three 128-dimension interaction layers. Other hyperparameters of the model are detailed in Chapter-S-IV in SI. We set the batch size to 32 and trained the model for 500 epochs. The variation of the loss function during model training is shown in Figure 2A, which shows that the training has converged. Meanwhile, the change in the loss function in the validation set indicates that there is no overfitting either. The performance of the trained MolCT-based force field on the test set is shown in Figure 3. The MolCT force field has a mean absolute error (MAE) of $0.2 \text{kcal} \cdot \text{mol}^{-1}$ for the energy and a root mean square error (RMSE) of $0.8 \text{ kcal} \cdot \text{mol}^{-1} \cdot \text{Å}^{-1}$ for the force, compared to the QM calculation. These results indicate that the MolCT force field has a high precision. In contrast, the errors in energy and force obtained by the DFTB method are much larger, as shown in Figure S2 in SI.

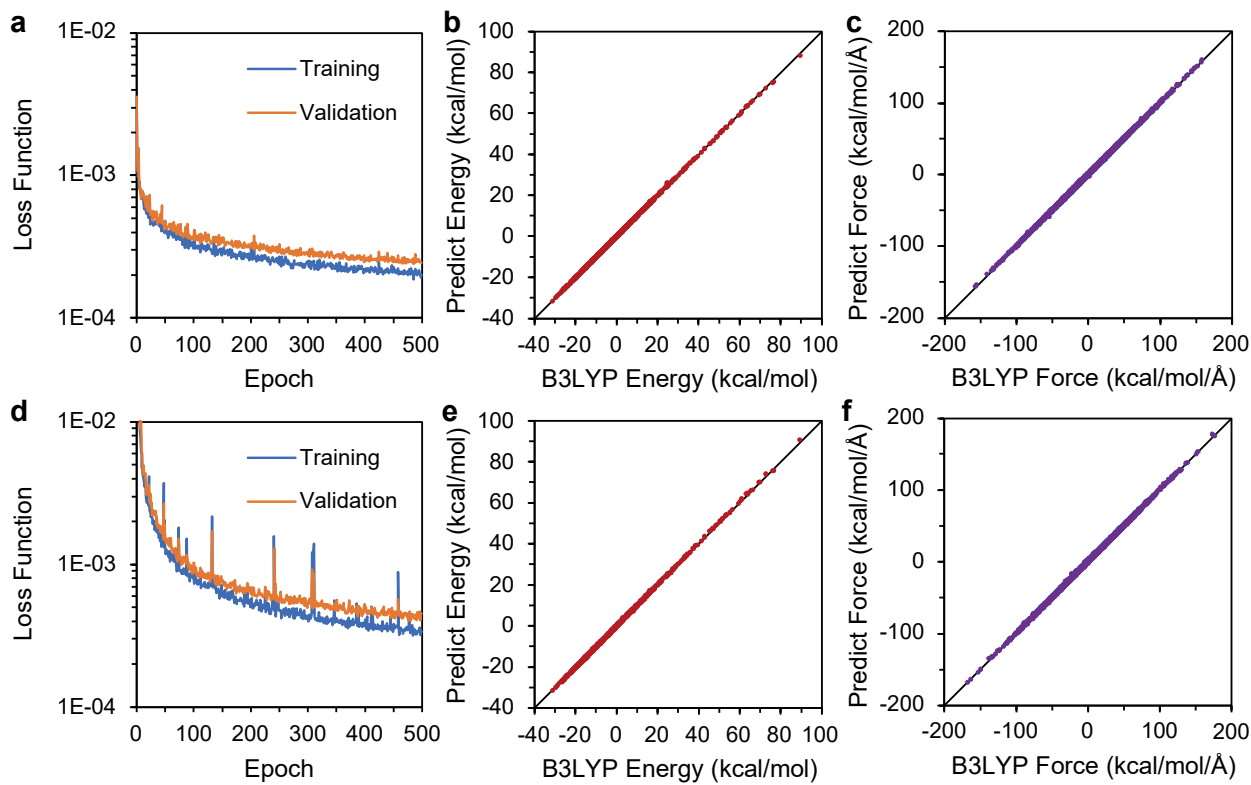

Figure 3. Results of two trainings of the MolCT force field for the Claisen rearrangement reaction, the graphs at the upper part are the results of the first training, and the graphs at the lower part are the results of the second training. a) and c) Changes of the loss function in the training and validation sets. b) and d) Performance of the energies predicted by the trained MolCT force field model on the test set. c) and e) Performance of the component forces (on different axes) predicted by the trained MolCT force field model on the test set.

Next, we used this MolCT force field to run the MD simulation. The setup of the MD simulation is the same as in the previous sampling of training data, and MW-MetaD enhanced sampling is used with the same CV. We obtained in parallel 100 trajectories in parallel, each for 300ps. For reasons discussed above and somehow expected, unreasonable conformations were found in the MD simulations on some of the trajectories. We randomly and uniformly selected 996 points along the CV to evaluate the performance of energy and force predictions during the dynamic simulations. Among them, 9 points did not converge during the SCF calculations, while several points showed abnormally high energies. These results showed that a supplement of the training dataset is necessary. We calculated the difference between MolCT energies and QM energies of the selected points, finding that the chemical structures of these molecular conformations with large errors are still reasonable, so we added all of them as supplementary conformations. In addition, we also performed trajectory shooting simulations, from which molecular conformations near the transition state were selected to complement the dataset (see Chapter-S-IV in SI for details). Ultimately, we selected a total of 7,734 conformations from the trajectories and used again B3LYP/6-31+G** method to compute their energies and forces as supplement to the training dataset. Subsequently, these additional 7,734 data points were included alongside the randomly selected half of the original dataset, resulting in 84,349 data points. This new dataset was partitioned into 83,197 data points for the training set, 128 for the validation set, and 1024 for the test set. Figure 3d illustrates the process of force field retraining. The performance of the retrained MolCT force field on the test set is then shown in Figure 3e and Figure 3f, with the MAE of energy as $0.23 \text{kcal} \cdot \text{mol}^{-1}$ and RMSE of force as $1.0 \text{ kcal} \cdot \text{mol}^{-1} \cdot \text{Å}^{-1}$.

We then performed MD simulations with the newly trained MolCT force field (See Movie S1) and found that it allows stable simulations for a long time. Thus, we performed 100 parallel MD simulations with WT-MetaD, each lasting 500ps, with the other parameters remaining the same as in the previous simulations. We calculated the free energy surface (FES) of the system as a function of CV based on the results of the MD simulations, as shown in Figure 4a. Using the weighted thermodynamic perturbation (wTP)[90] method, we also calculated the reweighted FES of the system using the B3LYP potential as a reference. In this procedure, QM calculations were performed to obtain the single point energy of the molecular conformation sampled under the MolCT force field. The difference between the FES of the MolCT force field and B3LYP is small, with a maximum of only 1.2 kcal/mol. As a comparison, we also calculated the FES based on the MD simulation with DFTB as the force field, which differs significantly from the MolCT force field calculation. Particularly, discrepancies were noted in the relative energies of Configuration A and B. The DFTB-derived FES suggests that Configuration A has lower free energy than Configuration B, contrary to the expected result where Configuration B should have lower free energy than Configuration A based on single point energy calculation and previous research of the reaction mechanism using SCC-DFTB method[16].

Furthermore, we also calculated the two-dimensional FES, which was plotted as a function of two distances, d1 and d2, for different force fields. Figures 4b and 4c show two FES calculated using the MolCT force field model and the DFTB, respectively. Significant differences can be found between the two diagrams, not only in the values of the free energies, but also in their overall landscapes. The landscape for MolCT forcefield shows a narrower channel from the saddle point to Configuration A than that obtained from DFTB method. DFTB calculation thus appears to provide an overestimate on the conformational flexibility of transition state. We then calculated the reweighted FES of B3LYP method using wTP and show the results in Figure 4d. We find that the reweighting of the MolCT FES to the B3LYP-level QM calculation brings almost no changes. This result is expected since the potential energy calculated using the MolCT force field is very close to that obtained by the QM calculation. Therefore, the accuracy of the MolCT force field for the chemical reaction is comparable to *ab initio* calculations and is much higher than that of the semi-empirical force fields.

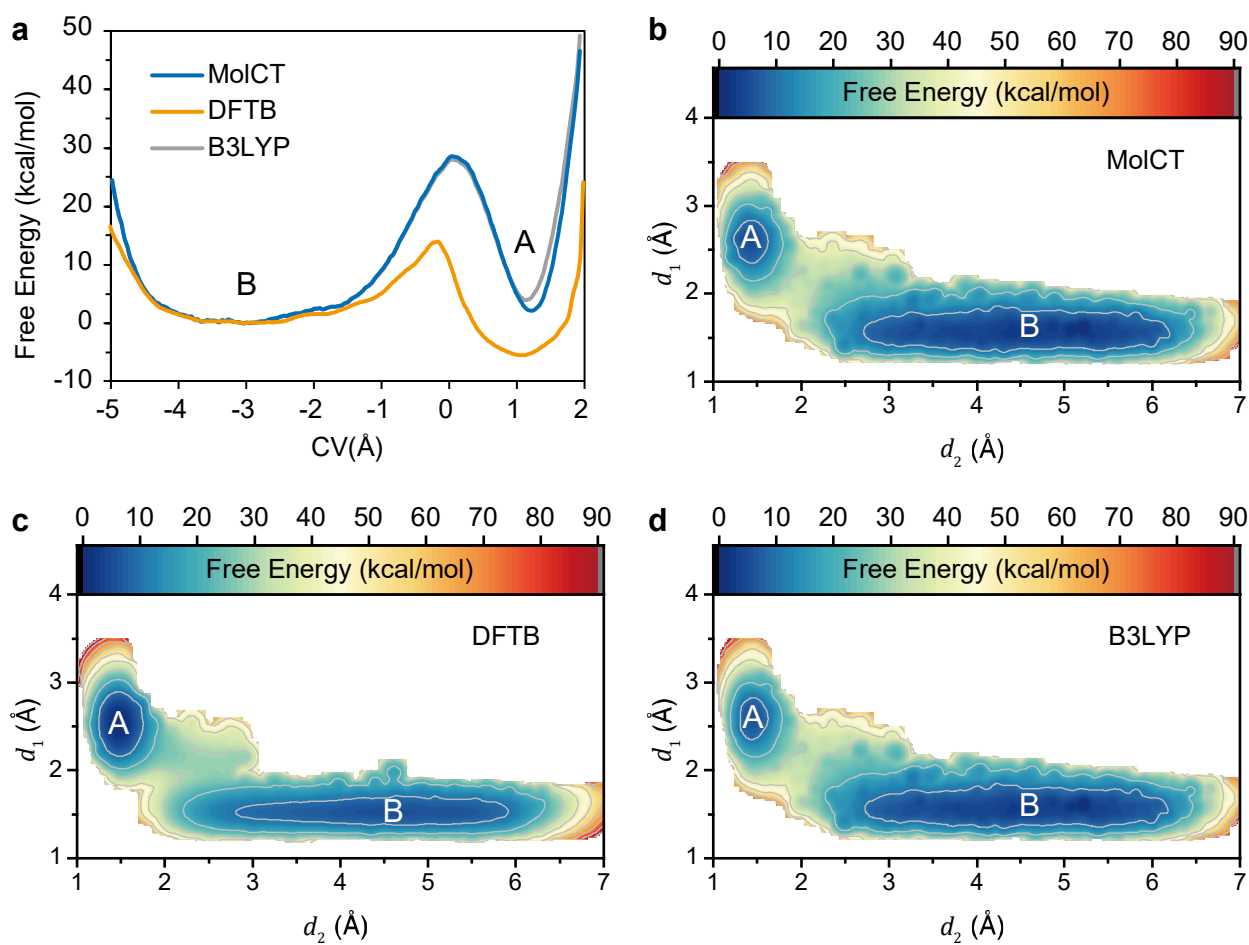

Figure 4 Free energy surfaces (FES) of the Claisen rearrangement reaction. a) FES as functions of CV under the MolCT force field, the DFTB method, and B3LYP method. b), c) and d) 2D FES as functions of distances $d_1$ and $d_2$ under the MolCT force field, the DFTB method, and B3LYP method, respectively.

**Carbonyl insertion reaction catalyzed by manganese**

We next tested our method on a more complex reaction, manganese-catalyzed carbonyl insertion. Alkyl migration reactions represent fundamental organometallic reactions involving transition metals. Figure 5a illustrates a simple example wherein $Mn(CO)_5CH_3$ reacts with carbon monoxide, resulting in the migration of a methyl group to a carbonyl moiety, producing $Mn(CO)_5(COCH_3)$. This reaction is also commonly referred to as carbonyl insertion[91] and consists of two steps[92]. Here we focused on the process of methyl group migration (Figure 5b).

We used the PM6[68] semi-empirical method, which supports the treatment of manganese ions, for MD simulation at 300 K. Due to the relative complexity of this reaction, we used the HLDA algorithm to find CVs for enhanced sampling, and two bond lengths, as depicted in Figure 5b, were chosen as descriptors. Results from the HLDA algorithm reveal a collective variable (CV) as CV = 0.99d1 + 0.14d2, which indicates that the impending formation of the carbon-carbon bond holds an overwhelmingly dominant role in this process. We obtained four parallel MD simulation trajectories for this system with MW-MetaD enhanced sampling methods. After 5 ns of simulation, the system has successfully transformed between reactants and products for multiple times. See Chapter-S-V in SI for simulation details.

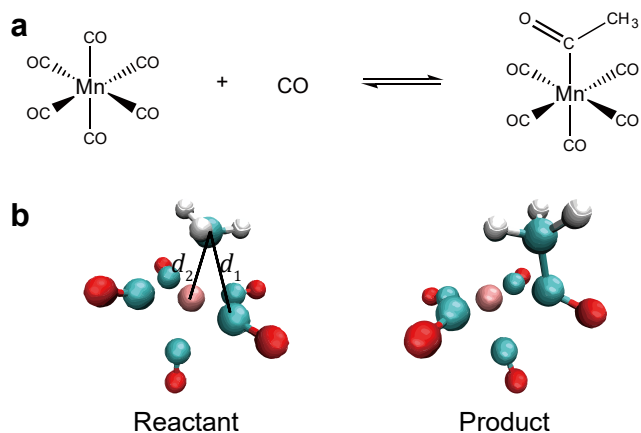

Figure 5. Carbonyl insertion reaction catalyzed by manganese. a) Chemical reaction formula. b) Snapshots of the 3D structures of the reactant and product, and the definition of the distances $d_1$ & $d_2$.

Next, we extracted the molecular conformations from the simulation trajectories to construct the training dataset. We categorized the molecular conformations with the range of $d_1$ between 1.73 to 1.87 Å as the transition states and retained them in the datasets. In the meantime, a random sampling was conducted for the reactant and product, resulting in a total of 133,120 samples. We calculated the single point energy and energy gradients for each molecular conformation using the B3LYP/6-311++G** method. We obtained a dataset of the atomic coordinates of these conformations along with the calculated energies and atomic forces and divide them randomly into three sets: 131,072 samples for the training set, 1,024 samples for the validation set, and another 1,024 samples for the test set.

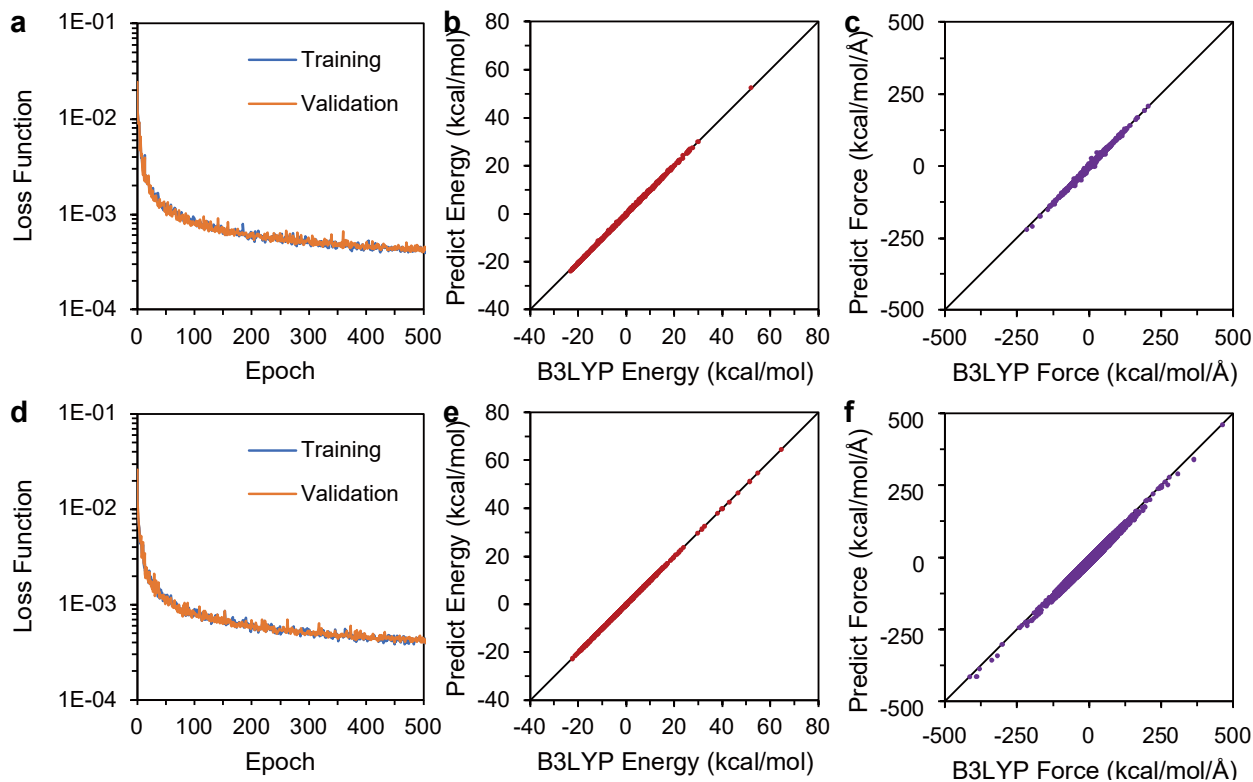

Figure 6. Results of two trainings of the MolCT force field for the carbonyl insertion reaction, the graphs at the upper part are the results of the first training, and the graphs at the lower part are the results of the second training. a) and c) Changes of the loss function in the training and validation sets. b) and d) Performance of the energies predicted by the trained MolCT force field model on the test set. c) and e) Performance of the component forces (on different axes) predicted by the trained MolCT force field model on the test set.

The dataset was then used to train the force field of the MolCT model using the same hyperparameters as discussed earlier. After 512 epochs of training, as depicted in Figure 6, the system's loss function has essentially converged. The performance of the trained MolCT model on the test set is shown in Figure 6. The MAE for energy has reduced to below 0.5 kcal/mol, while the RMSE for force is below 0.95 $\text{kcal} \cdot \text{mol}^{-1} \cdot \text{Å}^{-1}$, thus achieving chemical accuracy.

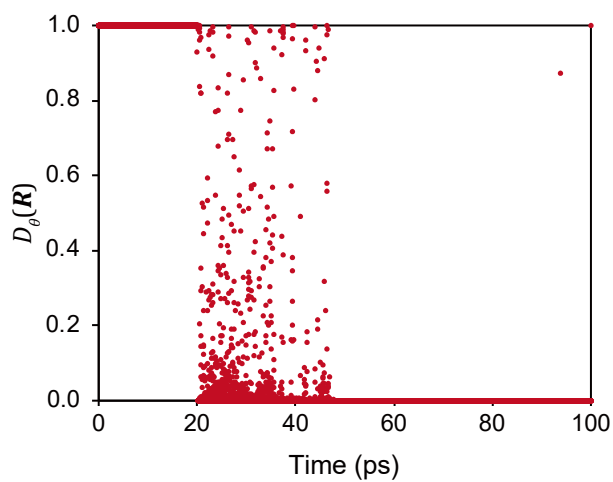

Figure 7. Evolution of the discriminator's value $D(\mathbf{R})$ for molecular conformation over time during the MD simulation of carbonyl insertion reaction with the first trained MolCT force field.

Following an initial run of MD simulation using the MolCT force field generated above, we supplemented the training dataset as described before. Because of the complexity of this chemical reaction and many irregular molecular structures appearing in the simulation trajectories, we train a discriminator $D_\theta(\mathbf{R})$ to supplement suitable conformations to the training dataset. The molecular conformations in the original training set were used as positive samples, while those conformations obtained from the MD simulations using MolCT as a force field were employed as negative samples. After training the discriminator $D_\theta(\mathbf{R})$ for 50 epoches using equation (10), we used it to analyze the MD simulated trajectories (Figure 7). The trained discriminator can accurately show the process by which the MD simulation becomes abnormal. The molecular conformations of the simulation trajectory in the initial stages have $D_\theta(\mathbf{R}) = 1$, indicating that they are identical to the conformations in the training set. In the later stages of the simulation trajectory, the $D_\theta(\mathbf{R})$ values of the molecular conformations begin to fluctuate and eventually become 0. This result indicates that conformations that do not exist in the training set begin to appear in the MD simulation and eventually turn into anomalous structures. Here, we chose molecular conformations of $0.2 < D < 0.999$ for DFT calculations. After excluding molecular conformations with unreasonably high energies, a total of 61,516 conformations were supplemented to the dataset (see Chapter-S-VI in SI for details). The evaluation of force field retraining on the test set is shown in Figure 6b. The retrained MolCT force field was tested to show stable MD simulations (See Movie S2). Therefore, we performed MD simulations of the manganese-catalyzed carbonyl insertion reaction with this latter force field using the MetaD enhanced sampling method. In 2 ns of simulation, the system was able to transform multiple times between reactants and products (Fig S6). Thus, we calculated the FES as a function of CV under the MolCT force field and compared with the FES under the semi-empirical method PM6. Similarly, using the wTP method, we calculated the reweighted FES of the system using the B3LYP method potential as a reference. As shown in Figure 8a, the FES under PM6 method deviates from the reference FES are extremely different, while the FES under MolCT force field is almost coincided with reference FES. In the MolCT force field, the free energy difference from the reactants to the products is 15.3 kcal/mol, while the free energy barrier from the reactants to the transition state is 15.6 kcal/mol. In contrast, under the PM6 force field, the free energy difference is only 0.2 kcal/mol and the energy barrier is 9.1 kcal/mol, which are far from the reference states and the reported values of 14.4 kcal/mol and an energy barrier of 14.5 kcal/mol[92]. Therefore, for more complex chemical reactions such as carbonyl insertion, the usual semi-empirical methods are no longer able to describe their thermodynamics correctly.

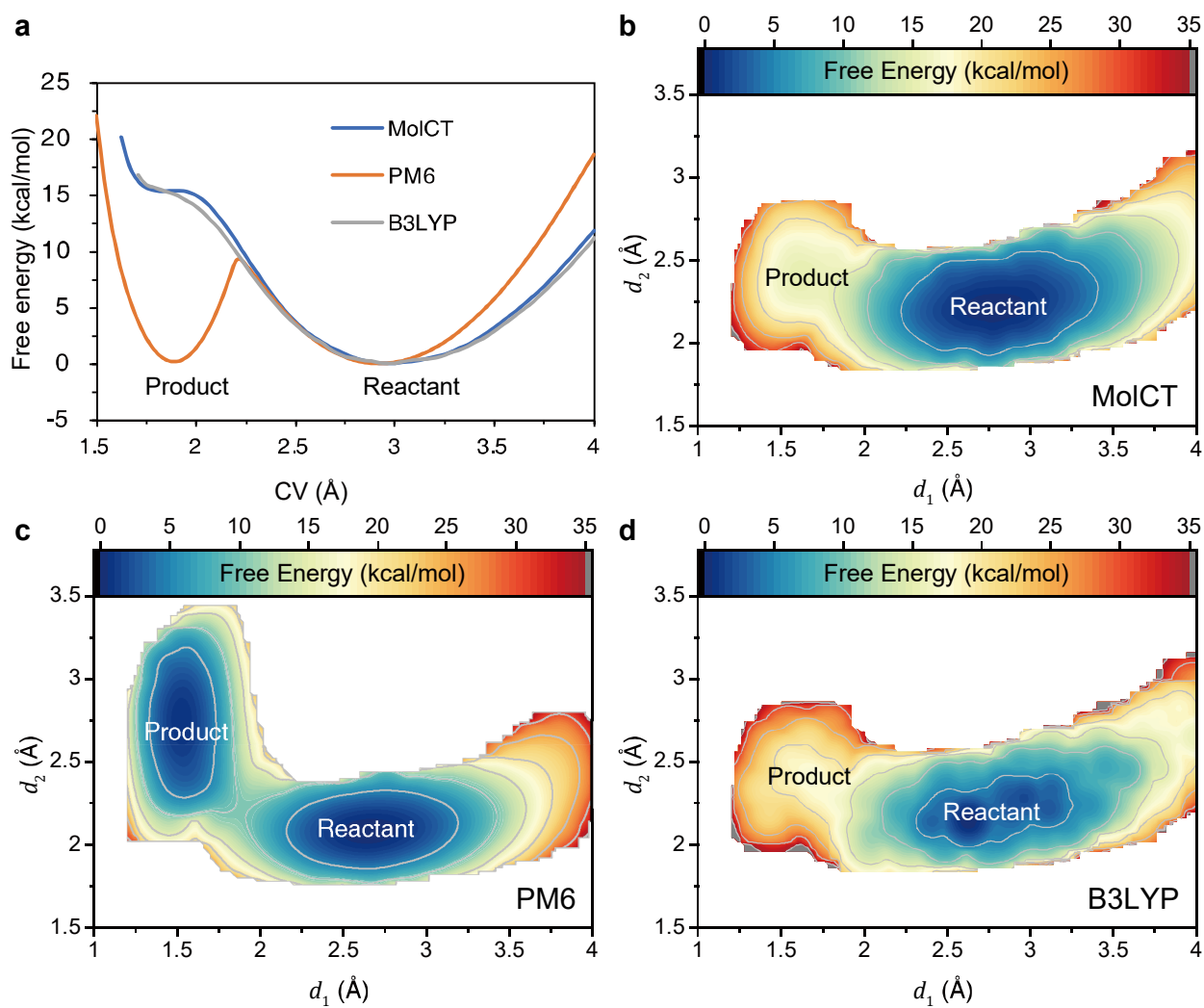

Figure 8. Free energy surfaces (FES) of the carbonyl insertion reaction. a) FES as functions of CV under the MolCT force field and the PM6 method. b) c) and d) 2D FES as functions of distances $d_1$ and $d_2$ under the MolCT force field, the PM6 method, and the reference, respectively.

Moreover, using the MolCT force field allows for a more accurate description of the mechanism of this reaction. We calculated the two-dimensional FES of the system as a function of d1 and d2 for two different force fields (Figure 8b and Figure 8c). We found that the landscape of the 2D-FES of this chemical reaction varies dramatically in different force fields. The FES of the MolCT force field in Figure 8b shows a larger conformational space for the reactants and a very small one for the products. In contrast, the FES of the PM6 force field in Figure 8c shows that the conformational space of the product is even larger than that of the reactant. More importantly, the positions and ranges of the transition states in the two FES are different, which would directly lead to a discrepancy in the description of the chemical reaction mechanism. We then calculated the reweighted FES of B3LYP method using wTP and show the results in Figure 8d. Again, we find that the reweighting of the MolCT FES to the B3LYP-level QM calculation brings small changes for the same reason that potential energy calculated using the MolCT force field is very close to that obtained by the QM calculation.

## Conclusion

In this paper, we propose an approach to train the deep molecular model MolCT as a high precision force field for MD simulations of chemical reactions. Benefiting from the high fitting accuracy with fewer network parameters of the MolCT model, the MolCT-based force field provides an efficient tool for chemical reaction simulations, with both high speed and accuracy. We have devised a scheme to efficiently generate datasets for force field training by combining enhanced sampling and multi-precision simulations. We also introduced discriminators from adversarial networks in the training process, which can most efficiently filter out molecular conformations lacking in the original dataset. This scheme produces highly accurate datasets that can cover the active region of the FES for chemical reactions at a small computational cost. This approach should also find use in AI and other areas of computational chemistry when the cost of obtaining dataset is high.

We used this approach to successfully generate high-precision force fields for MD simulations of two systems, the Claisen rearrangement and manganese carbonyl insertion reactions. Both systems need to be retrained only once to obtain force fields that allow stable and accurate MD simulations, demonstrating the efficiency of our strategy. The thermodynamic properties of the two systems calculated from MD simulations using the MolCT force field are more accurate compared to the semi-empirical approach. The FES from the MolCT force field is very close to the one calculated from the *ab initio* calculation, which differs considerably from the results of the semi-empirical method. For the more complex manganese-catalyzed carbonyl insertion reaction, the free energies calculated using semi-empirical methods differ greatly from the literature report and it does not explain the experimental mechanism. In contrast, the computational results using the MolCT force field are in good agreement with the experimental values.

We hope that this methodology can be used on other chemical reactions, such as the transition-metal-catalyzed organic reactions, which are important in biochemical studies[93] and industrial applications[64]. While current methods are limited to chemical reactions in vacuum or implicit solvent environments, we are developing new models that incorporate the effects of solvent and protein environments into the force field. This latter effort will further expand the scope of high-precision molecular dynamics studies.

## Acknowledgment

The authors thank Yijie Xia and Yanheng Li for useful discussion. Computational resources were supported by Shenzhen Bay Laboratory supercomputing center. This work was supported by National Science and Technology Major Project (No. 2022ZD0115003), National Natural Science Foundation of China (22003042, 22273061 to Y.I.Y., and 22050003, 92053202, 21821004, 21927901 to Y.Q.G.),

## Reference

(1) Cheng, G.-J.; Zhang, X.; Chung, L. W.; Xu, L.; Wu, Y.-D. Computational Organic Chemistry: Bridging Theory and Experiment in Establishing the Mechanisms of Chemical Reactions. *J. Am. Chem. Soc.* **2015**, *137* (5), 1706–1725. https://doi.org/10.1021/ja5112749.
(2) van Gunsteren, W. F.; Bakowies, D.; Baron, R.; Chandrasekhar, I.; Christen, M.; Daura, X.; Gee, P.; Geerke, D. P.; Glättli, A.; Hünenberger, P. H.; Kastenholz, M. A.; Oostenbrink, C.; Schenk, M.; Trzesniak, D.; van der Vegt, N. F. A.; Yu, H. B. Biomolecular Modeling: Goals, Problems, Perspectives. *Angewandte Chemie International Edition* **2006**, *45* (25), 4064–4092. https://doi.org/10.1002/anie.200502655.
(3) Friesner, R. A. Ab Initio Quantum Chemistry: Methodology And Applications. *Proc. Natl. Acad. Sci. U. S. A.* **2005**, *102*, 6648.
(4) Cheong, P. H.-Y.; Legault, C. Y.; Um, J. M.; Çelebi-Ölçüm, N.; Houk, K. N. Quantum Mechanical Investigations of Organocatalysis: Mechanisms, Reactivities, and Selectivities. *Chem. Rev.* **2011**, *111* (8), 5042–5137. https://doi.org/10.1021/cr100212h.
(5) Henkelman, G.; Uberuaga, B. P.; Jónsson, H. A Climbing Image Nudged Elastic Band Method for Finding Saddle Points and Minimum Energy Paths. *The Journal of chemical physics* **2000**, *113* (22), 9901–9904.
(6) Henkelman, G.; Jónsson, H. Improved Tangent Estimate in the Nudged Elastic Band Method for Finding Minimum Energy Paths and Saddle Points. *The Journal of chemical physics* **2000**, *113* (22), 9978–9985.
(7) Peters, B.; Heyden, A.; Bell, A. T.; Chakraborty, A. A Growing String Method for Determining Transition States: Comparison to the Nudged Elastic Band and String Methods. *The Journal of chemical physics* **2004**, *120* (17), 7877–7886.
(8) Weinan, E.; Ren, W.; Vanden-Eijnden, E. String Method for the Study of Rare Events. *Physical Review B* **2002**, *66* (5), 052301.
(9) Cerjan, C. J.; Miller, W. H. On Finding Transition States. *The Journal of chemical physics* **1981**, *75* (6), 2800–2806.
(10) Simons, J.; Joergensen, P.; Taylor, H.; Ozment, J. Walking on Potential Energy Surfaces. *The Journal of Physical Chemistry* **1983**, *87* (15), 2745–2753.

(11) Banerjee, A.; Adams, N.; Simons, J.; Shepard, R. Search for Stationary Points on Surfaces. *The Journal of Physical Chemistry* **1985**, *89* (1), 52–57.
(12) Heyden, A.; Bell, A. T.; Keil, F. J. Efficient Methods for Finding Transition States in Chemical Reactions: Comparison of Improved Dimer Method and Partitioned Rational Function Optimization Method. *The Journal of chemical physics* **2005**, *123* (22).
(13) Henkelman, G.; Jónsson, H. A Dimer Method for Finding Saddle Points on High Dimensional Potential Surfaces Using Only First Derivatives. *The Journal of chemical physics* **1999**, *111* (15), 7010–7022.
(14) Olsen, R. A.; Kroes, G. J.; Henkelman, G.; Arnaldsson, A.; Jónsson, H. Comparison of Methods for Finding Saddle Points without Knowledge of the Final States. *The Journal of chemical physics* **2004**, *121* (20), 9776–9792.
(15) Yang, Y.-F.; Cheng, G.-J.; Liu, P.; Leow, D.; Sun, T.-Y.; Chen, P.; Zhang, X.; Yu, J.-Q.; Wu, Y.-D.; Houk, K. N. Palladium-Catalyzed Meta-Selective C–H Bond Activation with a Nitrile-Containing Template: Computational Study on Mechanism and Origins of Selectivity. *Journal of the American Chemical Society* **2014**, *136* (1), 344–355.
(16) Zhang, J.; Yang, Y. I.; Yang, L.; Gao, Y. Q. Conformational Preadjustment in Aqueous Claisen Rearrangement Revealed by SITS-QM/MM MD Simulations. *J. Phys. Chem. B* **2015**, *119* (17), 5518–5530. https://doi.org/10.1021/jp511057f.
(17) Zhang, L.; An, K.; Wang, Y.; Wu, Y.-D.; Zhang, X.; Yu, Z.-X.; He, W. A Combined Computational and Experimental Study of Rh-Catalyzed C–H Silylation with Silacyclobutanes: Insights Leading to a More Efficient Catalyst System. *Journal of the American Chemical Society* **2021**, *143* (9), 3571–3582.
(18) Arcus, V. L.; van der Kamp, M. W.; Pudney, C. R.; Mulholland, A. J. Enzyme Evolution and the Temperature Dependence of Enzyme Catalysis. *Current Opinion in Structural Biology* **2020**, *65*, 96–101. https://doi.org/10.1016/j.sbi.2020.06.001.
(19) Sheu, S.-Y. Molecular Dynamics Simulation of Entropy Driven Ligand Escape Process in Heme Pocket. *The Journal of Chemical Physics* **2005**, *122* (10), 104905. https://doi.org/10.1063/1.1860552.
(20) Xie, L.; Yang, M.; Chen, Z.-N. Understanding the Entropic Effect in Chorismate Mutase Reaction Catalyzed by Isochorismate-Pyruvate Lyase from Pseudomonas Aeruginosa (PchB). *Catal. Sci. Technol.* **2019**, *9* (4), 957–965. https://doi.org/10.1039/C8CY02123F.
(21) Han, X.; Zhang, J.; Yang, Y. I.; Zhang, Z.; Yang, L.; Gao, Y. Q. Enhanced Sampling Simulation Reveals How Solvent Influences Chirogenesis of the Intra-Molecular Diels–Alder Reaction. *Journal of Chemical Theory and Computation* **2022**, *18* (7), 4318–4326.
(22) Spiwok, V.; Lipovová, P.; Králová, B. Metadynamics in Essential Coordinates: Free Energy Simulation of Conformational Changes. *J. Phys. Chem. B* **2007**, *111* (12), 3073–3076. https://doi.org/10.1021/jp068587c.
(23) Tavernelli, I.; Cotesta, S.; Di Iorio, E. E. Protein Dynamics, Thermal Stability, and Free-Energy Landscapes: A Molecular Dynamics Investigation. *Biophysical Journal* **2003**, *85* (4), 2641–2649. https://doi.org/10.1016/S0006-3495(03)74687-6.
(24) Mondal, M.; Yang, L.; Cai, Z.; Patra, P.; Gao, Y. Q. A Perspective on the Molecular Simulation of DNA from Structural and Functional Aspects. *Chemical Science* **2021**, *12* (15), 5390–5409.
(25) Baker, C. M.; Best, R. B. Insights into the Binding of Intrinsically Disordered Proteins from Molecular Dynamics Simulation. *WILEY INTERDISCIPLINARY REVIEWS-COMPUTATIONAL MOLECULAR SCIENCE* **2014**, *4* (3), 182–198. https://doi.org/10.1002/wcms.1167.
(26) Lei, Y. K.; Zhang, J.; Zhang, Z.; Gao, Y. Q. Dynamic Electric Field Complicates Chemical Reactions in Solutions. *The Journal of Physical Chemistry Letters* **2019**, *10* (11), 2991–2997.
(27) Yang, Y. I.; Shao, Q.; Zhang, J.; Yang, L.; Gao, Y. Q. Enhanced Sampling in Molecular Dynamics. *The Journal of chemical physics* **2019**, *151* (7).
(28) Laio, A.; Parrinello, M. Escaping Free-Energy Minima. *Proceedings of the national academy of sciences* **2002**, *99* (20), 12562–12566.
(29) Valsson, O.; Tiwary, P.; Parrinello, M. Enhancing Important Fluctuations: Rare Events and

Metadynamics from a Conceptual Viewpoint. *Annual review of physical chemistry* **2016**, *67*, 159–184.

(30) Ponte, F.; Piccini, G.; Sicilia, E.; Parrinello, M. A Metadynamics Perspective on the Reduction Mechanism of the Pt (IV) Asplatin Prodrug. *Journal of Computational Chemistry* **2020**, *41* (4), 290–294.

(31) Danese, M.; Bon, M.; Piccini, G.; Passerone, D. The Reaction Mechanism of the Azide–Alkyne Huisgen Cycloaddition. *Physical Chemistry Chemical Physics* **2019**, *21* (35), 19281–19287.

(32) Adda, A.; Aoul, R. H.; Sediki, H.; Sehailia, M.; Krallafa, A. M. Selectivity in the Wittig Reaction within the Ab Initio Static and Metadynamics Approaches. *Theor Chem Acc* **2023**, *142* (10), 102. https://doi.org/10.1007/s00214-023-03029-1.

(33) Gao, Y. Q. An Integrate-over-Temperature Approach for Enhanced Sampling. *The Journal of chemical physics* **2008**, *128* (6).

(34) Yang, L.; Liu, C.-W.; Shao, Q.; Zhang, J.; Gao, Y. Q. From Thermodynamics to Kinetics: Enhanced Sampling of Rare Events. *Acc. Chem. Res.* **2015**, *48* (4), 947–955. https://doi.org/10.1021/ar500267n.

(35) Shao, Q.; Gao, Y. Q. Temperature Dependence of Hydrogen-Bond Stability in β-Hairpin Structures. *J. Chem. Theory Comput.* **2010**, *6* (12), 3750–3760. https://doi.org/10.1021/ct100436r.

(36) Shao, Q.; Wei, H.; Gao, Y. Q. Effects of Turn Stability and Side-Chain Hydrophobicity on the Folding of β-Structures. *Journal of Molecular Biology* **2010**, *402* (3), 595–609. https://doi.org/10.1016/j.jmb.2010.08.037.

(37) Zhang, J.; Yang, Y. I.; Yang, L.; Gao, Y. Q. Dynamics and Kinetics Study of "In-Water" Chemical Reactions by Enhanced Sampling of Reactive Trajectories. *J. Phys. Chem. B* **2015**, *119* (45), 14505–14514. https://doi.org/10.1021/acs.jpcb.5b08690.

(38) Liu, C.-W.; Wang, F.; Yang, L.; Li, X.-Z.; Zheng, W.-J.; Gao, Y. Q. Stable Salt–Water Cluster Structures Reflect the Delicate Competition between Ion–Water and Water–Water Interactions. *J. Phys. Chem. B* **2014**, *118* (3), 743–751. https://doi.org/10.1021/jp408439j.

(39) Hou, G.-L.; Liu, C.-W.; Li, R.-Z.; Xu, H.-G.; Gao, Y. Q.; Zheng, W.-J. Emergence of Solvent-Separated Na+–Cl– Ion Pair in Salt Water: Photoelectron Spectroscopy and Theoretical Calculations. *J. Phys. Chem. Lett.* **2017**, *8* (1), 13–20. https://doi.org/10.1021/acs.jpclett.6b02670.

(40) He, Z.; Feng, G.; Yang, B.; Yang, L.; Liu, C.-W.; Xu, H.-G.; Xu, X.-L.; Zheng, W.-J.; Gao, Y. Q. Molecular Dynamics Simulation, Ab Initio Calculation, and Size-Selected Anion Photoelectron Spectroscopy Study of Initial Hydration Processes of Calcium Chloride. *The Journal of Chemical Physics* **2018**, *148* (22), 222839. https://doi.org/10.1063/1.5024279.

(41) Yang, Y. I.; Zhang, J.; Che, X.; Yang, L.; Gao, Y. Q. Efficient Sampling over Rough Energy Landscapes with High Barriers: A Combination of Metadynamics with Integrated Tempering Sampling. *The Journal of Chemical Physics* **2016**, *144* (9).

(42) Yang, Y. I.; Niu, H.; Parrinello, M. Combining Metadynamics and Integrated Tempering Sampling. *The Journal of Physical Chemistry Letters* **2018**, *9* (22), 6426–6430.

(43) Mendels, D.; Piccini, G.; Parrinello, M. Collective Variables from Local Fluctuations. *The journal of physical chemistry letters* **2018**, *9* (11), 2776–2781.

(44) Piccini, G.; Mendels, D.; Parrinello, M. Metadynamics with Discriminants: A Tool for Understanding Chemistry. *Journal of chemical theory and computation* **2018**, *14* (10), 5040–5044.

(45) Schwantes, C. R.; Pande, V. S. Improvements in Markov State Model Construction Reveal Many Non-Native Interactions in the Folding of NTL9. *J. Chem. Theory Comput.* **2013**, *9* (4), 2000–2009. https://doi.org/10.1021/ct300878a.

(46) Pérez-Hernández, G.; Paul, F.; Giorgino, T.; De Fabritiis, G.; Noé, F. Identification of Slow Molecular Order Parameters for Markov Model Construction. *The Journal of Chemical Physics* **2013**, *139* (1), 015102. https://doi.org/10.1063/1.4811489.

(47) MacFadyen, J.; Andricioaei, I. A Skewed-Momenta Method to Efficiently Generate Conformational-Transition Trajectories. *The Journal of chemical physics* **2005**, *123* (7).

(48) Tiwary, P.; Berne, B. J. Spectral Gap Optimization of Order Parameters for Sampling Complex Molecular Systems. *Proceedings of the National Academy of Sciences* **2016**, *113* (11), 2839–2844.

(49) Elstner, M. The SCC-DFTB Method and Its Application to Biological Systems. *Theoretical Chemistry Accounts* **2006**, *116*, 316–325.
(50) Stewart, J. J. Application of the PM6 Method to Modeling Proteins. *Journal of molecular modeling* **2009**, *15* (7), 765–805.
(51) Dewar, M. J.; Zoebisch, E. G.; Healy, E. F.; Stewart, J. J. Development and Use of Quantum Mechanical Molecular Models. 76. AM1: A New General Purpose Quantum Mechanical Molecular Model. *Journal of the American Chemical Society* **1985**, *107* (13), 3902–3909.
(52) Van Duin, A. C.; Dasgupta, S.; Lorant, F.; Goddard, W. A. ReaxFF: A Reactive Force Field For Hydrocarbons. *J. Phys. Chem. A* **2001**, *105*, 9396.
(53) Senftle, T. P.; Hong, S.; Islam, M. M.; Kylasa, S. B.; Zheng, Y.; Shin, Y. K.; Junkermeier, C.; Engel-Herbert, R.; Janik, M. J.; Aktulga, H. M.; Verstraelen, T.; Grama, A.; van Duin, A. C. T. The ReaxFF Reactive Force-Field: Development, Applications and Future Directions. *npj Comput Mater* **2016**, *2* (1), 1–14. https://doi.org/10.1038/npjcompumats.2015.11.
(54) Behler, J.; Parrinello, M. Generalized Neural-Network Representation of High-Dimensional Potential-Energy Surfaces. *Phys. Rev. Lett.* **2007**, *98*, 146401.
(55) Zhang, L.; Han, J.; Wang, H.; Car, R.; Weinan, E. Deep Potential Molecular Dynamics: A Scalable Model With The Accuracy Of Quantum Mechanics. *Phys. Rev. Lett.* **2018**, *120*, 143001.
(56) Zhang, Y.; Hu, C.; Jiang, B. Embedded Atom Neural Network Potentials: Efficient And Accurate Machine Learning With A Physically Inspired Representation. *J. Phys. Chem. Lett.* **2019**, *10*, 4962.
(57) Schütt, K. T.; Sauceda, H. E.; Kindermans, P.-J.; Tkatchenko, A.; Müller, K.-R. SchNet – A Deep Learning Architecture For Molecules And Materials. *J. Chem. Phys.* **2018**, *148*, 241722.
(58) Unke, O. T.; Meuwly, M. PhysNet: A Neural Network for Predicting Energies, Forces, Dipole Moments, and Partial Charges. *J. Chem. Theory Comput.* **2019**, *15*, 3678.
(59) Riedmiller, K.; Reiser, P.; Bobkova, E.; Maltsev, K.; Gryn'ova, G.; Friederich, P.; Gräter, F. Substituting Density Functional Theory in Reaction Barrier Calculations for Hydrogen Atom Transfer in Proteins. *Chem. Sci.* **2024**. https://doi.org/10.1039/D3SC03922F.
(60) Pan, X.; Snyder, R.; Wang, J.-N.; Lander, C.; Wickizer, C.; Van, R.; Chesney, A.; Xue, Y.; Mao, Y.; Mei, Y.; Pu, J.; Shao, Y. Training Machine Learning Potentials for Reactive Systems: A Colab Tutorial on Basic Models. *Journal of Computational Chemistry* *n/a* (n/a). https://doi.org/10.1002/jcc.27269.
(61) Zhang, J.; Zhou, Y.; Lei, Y.-K.; Yang, Y. I.; Gao, Y. Q. *Molecular CT: Unifying Geometry and Representation Learning for Molecules at Different Scales*. arXiv.org. https://arxiv.org/abs/2012.11816v2 (accessed 2023-10-10).
(62) Scarselli, F.; Gori, M.; Tsoi, A. C.; Hagenbuchner, M.; Monfardini, G. The Graph Neural Network Model. *IEEE Trans. Neural Netw.* **2009**, *20*, 61.
(63) Vaswani, A.; Shazeer, N.; Parmar, N.; Uszkoreit, J.; Jones, L.; Gomez, A. N.; Kaiser, L.; Polosukhin, I. Attention Is All You Need. *Advances in Neural Information Processing Systems* **2017**, 5998.
(64) Han, X.; Sun, T.-Y.; Yang, Y. I.; Zhang, J.; Qiu, J.; Xiong, Z.; Qiao, N.; Wu, Y.-D.; Gao, Y. Q. Enhanced Sampling Simulations on Transition-Metal-Catalyzed Organic Reactions: Zirconocene-Catalyzed Propylene Polymerization and Sharpless Epoxidation. *CCS Chemistry* **2023**, *0* (0), 1–12. https://doi.org/10.31635/ccschem.023.202302901.
(65) Porezag, D.; Frauenheim, T.; Köhler, T.; Seifert, G.; Kaschner, R. Construction of Tight-Binding-like Potentials on the Basis of Density-Functional Theory: Application to Carbon. *Physical Review B* **1995**, *51* (19), 12947.
(66) Seifert, G.; Porezag, D.; Frauenheim, T. Calculations of Molecules, Clusters, and Solids with a Simplified LCAO-DFT-LDA Scheme. *International journal of quantum chemistry* **1996**, *58* (2), 185–192.
(67) Seabra, G. de M.; Walker, R. C.; Elstner, M.; Case, D. A.; Roitberg, A. E. Implementation of the SCC-DFTB Method for Hybrid QM/MM Simulations within the Amber Molecular Dynamics Package. *The Journal of Physical Chemistry A* **2007**, *111* (26), 5655–5664.
(68) Stewart, J. J. Optimization of Parameters for Semiempirical Methods V: Modification of NDDO Approximations and Application to 70 Elements. *Journal of Molecular modeling* **2007**, *13*, 1173–

1213.

(69) Goodfellow, I.; Pouget-Abadie, J.; Mirza, M.; Xu, B.; Warde-Farley, D.; Ozair, S.; Courville, A.; Bengio, Y. Generative Adversarial Nets. *Advances in Neural Information Processing Systems* **2014**, 2672.

(70) Case, D. A.; Belfon, K.; Ben-Shalom, I. Y.; Brozell, S. R.; Cerutti, D. S.; Cheatham III, T. E.; Cruzeiro, V. W. D.; Darden, T. A.; Duke, R. E.; Giambasu, G. AMBER 20. 2020. *Google Scholar There is no corresponding record for this reference*.

(71) Brooks, B. R.; Brooks III, C. L.; Mackerell Jr, A. D.; Nilsson, L.; Petrella, R. J.; Roux, B.; Won, Y.; Archontis, G.; Bartels, C.; Boresch, S. CHARMM: The Biomolecular Simulation Program. *Journal of computational chemistry* **2009**, *30* (10), 1545–1614.

(72) Brooks, B. R.; Bruccoleri, R. E.; Olafson, B. D.; States, D. J.; Swaminathan, S. a; Karplus, M. CHARMM: A Program for Macromolecular Energy, Minimization, and Dynamics Calculations. *Journal of computational chemistry* **1983**, *4* (2), 187–217.

(73) Kühne, T. D.; Iannuzzi, M.; Del Ben, M.; Rybkin, V. V.; Seewald, P.; Stein, F.; Laino, T.; Khaliullin, R. Z.; Schütt, O.; Schiffmann, F. CP2K: An Electronic Structure and Molecular Dynamics Software Package-Quickstep: Efficient and Accurate Electronic Structure Calculations. *The Journal of Chemical Physics* **2020**, *152* (19).

(74) MindSPONGE: Simlation Package Towards Next Generation Molecular Modelling, 2021. https://gitee.com/mindspore/mindscience/tree/master/MindSPONGE.

(75) Zhang, J.; Chen, D.; Xia, Y.; Huang, Y.-P.; Lin, X.; Han, X.; Ni, N.; Wang, Z.; Yu, F.; Yang, L.; Yang, Y. I.; Gao, Y. Q. Artificial Intelligence Enhanced Molecular Simulations. *J. Chem. Theory Comput.* **2023**, *19* (14), 4338–4350. https://doi.org/10.1021/acs.jctc.3c00214.

(76) Sun, Q.; Berkelbach, T. C.; Blunt, N. S.; Booth, G. H.; Guo, S.; Li, Z.; Liu, J.; McClain, J. D.; Sayfutyarova, E. R.; Sharma, S. PySCF: The Python-based Simulations of Chemistry Framework. *Wiley Interdisciplinary Reviews: Computational Molecular Science* **2018**, *8* (1), e1340.

(77) Bonomi, M.; Bussi, G.; Camilloni, C.; Tribello, G. A.; Banáš, P.; Barducci, A.; Bernetti, M.; Bolhuis, P. G.; Bottaro, S.; Branduardi, D. Promoting Transparency and Reproducibility in Enhanced Molecular Simulations. *Nature methods* **2019**, *16* (8), 670–673.

(78) Tribello, G. A.; Bonomi, M.; Branduardi, D.; Camilloni, C.; Bussi, G. PLUMED 2: New Feathers for an Old Bird. *Computer physics communications* **2014**, *185* (2), 604–613.

(79) Bonomi, M.; Branduardi, D.; Bussi, G.; Camilloni, C.; Provasi, D.; Raiteri, P.; Donadio, D.; Marinelli, F.; Pietrucci, F.; Broglia, R. A. PLUMED: A Portable Plugin for Free-Energy Calculations with Molecular Dynamics. *Computer Physics Communications* **2009**, *180* (10), 1961–1972.

(80) Fiorin, G.; Klein, M. L.; Hénin, J. Using Collective Variables to Drive Molecular Dynamics Simulations. *Molecular Physics* **2013**, *111* (22–23), 3345–3362.

(81) Frisch, M. ea; Trucks, G. W.; Schlegel, H. B.; Scuseria, G. E.; Robb, M. A.; Cheeseman, J. R.; Scalmani, G.; Barone, V.; Petersson, G. A.; Nakatsuji, H. *Gaussian 16, Revision C. 01*; Gaussian, Inc., Wallingford CT, 2016.

(82) Neese, F. The ORCA Program System. *Wiley Interdisciplinary Reviews: Computational Molecular Science* **2012**, *2* (1), 73–78.

(83) Neese, F. Software Update: The ORCA Program System, Version 4.0. *Wiley Interdisciplinary Reviews: Computational Molecular Science* **2018**, *8* (1), e1327.

(84) Neese, F.; Wennmohs, F.; Becker, U.; Riplinger, C. The ORCA Quantum Chemistry Program Package. *The Journal of chemical physics* **2020**, *152* (22).

(85) Raiteri, P.; Laio, A.; Gervasio, F. L.; Micheletti, C.; Parrinello, M. Efficient Reconstruction of Complex Free Energy Landscapes by Multiple Walkers Metadynamics. *The journal of physical chemistry B* **2006**, *110* (8), 3533–3539.

(86) Becke, A. D. Density-functional Thermochemistry. I. The Effect of the Exchange-only Gradient Correction. *The Journal of chemical physics* **1992**, *96* (3), 2155–2160.

(87) Lee, C.; Yang, W.; Parr, R. G. Accurate and Simple Analytic Representation of the Electron-Gas Correlation Energy. *Phys. Rev. B* **1988**, *37*, 785–789.

(88) Vosko, S. H.; Wilk, L.; Nusair, M. Accurate Spin-Dependent Electron Liquid Correlation Energies for Local Spin Density Calculations: A Critical Analysis. *Canadian Journal of physics* **1980**, *58* (8), 1200–1211.
(89) Stephens, P. J.; Devlin, F. J.; Chabalowski, C. F.; Frisch, M. J. Ab Initio Calculation of Vibrational Absorption and Circular Dichroism Spectra Using Density Functional Force Fields. *The Journal of physical chemistry* **1994**, *98* (45), 11623–11627.
(90) Li, P.; Jia, X.; Pan, X.; Shao, Y.; Mei, Y. Accelerated Computation of Free Energy Profile at Ab Initio Quantum Mechanical/Molecular Mechanics Accuracy via a Semi-Empirical Reference Potential. I. Weighted Thermodynamics Perturbation. *J. Chem. Theory Comput.* **2018**, *14* (11), 5583–5596. https://doi.org/10.1021/acs.jctc.8b00571.
(91) Tanjaroon, C.; Zhou, Z.; Mills, D.; Keck, K.; Kukolich, S. G. Microwave Spectra and Theoretical Calculations for Two Structural Isomers of Methylmanganese Pentacarbonyl. *Inorganic chemistry* **2020**, *59* (9), 6432–6438.
(92) Derecskei-Kovacs, A.; Marynick, D. S. A New Look at an Old Reaction: The Potential Energy Surface for the Thermal Carbonylation of Mn (CO) 5CH3. The Role of Two Energetically Competitive Intermediates on the Reaction Surface, and Comments on the Photodecarbonylation of Mn (CO) 5 (COCH3). *Journal of the American Chemical Society* **2000**, *122* (9), 2078–2086.
(93) Wu, R.; Xie, H.; Cao, Z.; Mo, Y. Combined Quantum Mechanics/Molecular Mechanics Study on the Reversible Isomerization of Glucose and Fructose Catalyzed by Pyrococcus Furiosus Phosphoglucose Isomerase. *J. Am. Chem. Soc.* **2008**, *130* (22), 7022–7031. https://doi.org/10.1021/ja710633c.